\documentclass[webversion]{usetex-v1}

\usepackage{cite}	
\usepackage{url}
\usepackage{xspace}
\usepackage{graphicx}

\long\def\com#1{}

\begin{document}

\title{\vspace{-1em}
	Plugging Side-Channel Leaks with Timing Information Flow Control}

\author{
	\authname{Bryan Ford}
	\authaddr{Yale University}
	\authurl{\url{http://bford.info/}}
	\vspace{-2em}
}

\maketitle

\begin{abstract}

The cloud model's dependence on
massive parallelism and resource sharing
exacerbates the security challenge of timing side-channels.
Timing Information Flow Control (TIFC) is
a novel adaptation of IFC techniques
that may offer a way to reason about, and ultimately control,
the flow of sensitive information through systems
via timing channels.
With TIFC, objects such as files, messages, and processes
carry not just {\em content labels} describing the ownership
of the object's ``bits,''
but also {\em timing labels} describing information
contained in timing events affecting the object,
such as process creation/termination or message reception.
With two system design tools---%
{\em deterministic execution} and {\em pacing queues}---%
TIFC enables the construction
of ``timing-hardened'' cloud infrastructure
that permits statistical multiplexing,
while aggregating and rate-limiting timing information leakage
between hosted computations.

\end{abstract}

\section{Introduction}
\label{sec-intro}

Timing channels have been known and studied for decades~\cite{
	kemmerer83shared,wray91analysis},
but have received a resurgence of attention,
particularly in the cloud computing
context~\cite{ford10determinating,ristenpart09cloud}.
While decentralized information flow mechanisms~\cite{
	myers97decentralized,zeldovich06making}
and trusted computing hardware and security kernels~\cite{
	keller10nohype,zhang11cloudvisor}
may protect cloud data and computations
from break-ins and software vulnerabilities,
these mechanisms offer no defense against
information leakage via timing side-channels,
which are abundant in massively parallel environments.
Information theft may be possible even without active malware infection
of a ``victim'' computation,
as demonstrated in proof-of-concept via
shared L1 data cache~\cite{percival05cache},
shared functional units~\cite{wang06covert},
branch target cache~\cite{aciicmez07predicting}, and
instruction cache~\cite{aciicmez07yet}.

Although timing channels may represent a security risk
in any shared infrastructure,
the cloud model exacerbates these risks
in at least four ways,
discussed in more detail elsewhere~\cite{ford10determinating}
and briefly summarized here.

First,
	{\em parallelism makes timing channel pervasive.}
	Many processing resources yield timing channels,
	and even if one channel is exploitable only at low rate,
	an attacker who can gain co-residency with a victim
	on multiple nodes and cores in a cloud~\cite{ristenpart09cloud}
	can multiply the leakage rate by the level of parallelism.

Second,
	{\em insider attacks become outsider attacks.}
	An attacker would have to break into
	or get an account on private computing infrastructure
	before mounting a timing attack,
	but on the cloud the attacker need only
	purchase the necessary CPU time from the provider.

Third,
	{\em cloud-based timing attacks are unlikely to be caught.}
	Timing attacks
	do not violate conventional system protection invariants,
	and are unlikely to set off alarms or leave a trail of evidence.
	Further, while the owner of a private machine
	can scan running computations for malicious activity,
	a cloud provider has no prerogative to monitor its customers,
	and ironically,
	by doing so could invite privacy concerns or lawsuits.

Fourth,
	{\em the cloud business model depends on statistical multiplexing.}
	The classic approach to limit timing channels,
	``reserving'' hardware resources or timeslices to customers
	in a demand-independent fashion,
	would prevent the provider from oversubscribing
	and statistically multiplexing hardware resources for efficiency.

This paper introduces and informally explores an extension
of decentralized information flow control (DIFC) techniques~\cite{
	myers97decentralized,zeldovich06making}
to the task of reasoning about and controlling timing side-channels
between computations hosted in a provider's cloud infrastructure.
The approach currently addresses only timing side-channels
{\em internal} to a cloud and not, for example,
resulting from communication patterns visible on a public network~\cite{
	brumley03remote}.
Furthermore, this paper merely explores one potential approach,
which has not been rigorously formalized or experimentally validated;
doing so remains future work.


\section{Timing Information Flow Control}
\label{sec-tifc}

This section introduces
{\em Timing Information Flow Control} or TIFC,
an extension of DIFC
for reasoning about and control the propagation of sensitive information
into, out of, or within a software system via timing channels.
With TIFC,
an operating system can attach explicit labels or {\em taints}
to processes and other objects,
describing what sources, types, and granularities of timing information
may have affected the state of the labeled object.
Using these labels,
the OS can enforce policies constraining
how timing-derived information may flow among processes
and affect their results.


\subsection{TIFC Model Overview}

Our TIFC model
builds on Flume~\cite{
        krohn07information},
due to its simplicity and elegance.
\com{
Comparable TIFC models could probably be built ``stand-alone''
or as extensions to other DIFC systems, however.
Like Flume, our model
uses {\em tags} and {\em labels} to track information
as it flows through a system---%
potentially any type of information,
but here we focus exclusively on timing information.

}%
As in Flume,
we assign {\em labels} to system objects
such as processes, messages, and files.
A label can contain any number of {\em tags},
each of indicating that the labelled object has a particular ``taint,''
or may be derived from information owned by a particular user.
Unlike conventional DIFC, however,
TIFC labels reflect not only the {\em content} contained in such an object---%
i.e., the information contained in the bits
comprising a message or a process's state---%
but also information that may have affected the timing
of observable {\em events} associated with that object---%
a process starting or stopping, a message being sent or received, etc.
Consistent with conventional, informal practices
for reasoning about timing channels~\cite{kemmerer83shared,wray91analysis},
our TIFC model does not attempt the likely-infeasible task
of eliminating timing channels entirely,
but rather seeks to impose limits on the {\em rate}
at which information might leak via timing channels.

To distinguish content and timing taint explicitly,
we give TIFC labels the form $\{L_C/L_T\}$,
where $L_C$ is a set of tags representing content taint,
and $L_T$ is a set of tags representing timing taint.
As in Flume, content tags in the set $L_C$
simply identify a user, such as Alice or Bob.
Timing tags, however, we give the form $P_f$,
where $U$ is a user such as Alice or Bob,
and $f$ is a frequency representing the maximum rate
with which user $U$'s information might leak via this timing event,
in bits per second.
The frequency part of a timing tag may be $\infty$,
indicating that information leakage may occur at an unbounded rate.
Thus, the label $\{A/A_\infty,B_f\}$ attached to a message
might indicate that the content (bits) comprising the message
contains Alice's (and only Alice's) information,
but that the {\em timing} with which the message was sent
might contain (and hence leak) both Alice's and Bob's information---%
at an arbitrarily high rate in Alice's case,
but up to at most $f$ bits per second in Bob's case.

\subsubsection{Declassification Capabilities}

To enforce information security policies,
we similarly build on Flume's model.
We allow a process $P$ to transmit information
to another process or target object $O$
only if $P$'s label is a subset of $O$'s,
or if $P$ holds {\em declassification capabilities}
for any tags in $P$ that are not in $O$.
A {\em content declassification capability} has the form $U^-$,
and represents the ability to remove content tag $U$,
as in Flume.
TIFC also adds {\em timing declassification capabilities}
of the form $U^-_f$,
representing the ability to declassify information
carried by timing channels, at a rate up to frequency $f$.
The ``maximum-strength'' timing declassifier
$U^-_\infty$ is equivalent
to the content declassifier $U^-$;
timing capabilities with finite frequencies
represent weakened versions of these infinite-rate capabilities.

Suppose process $P_1$ has label $\{A/A_\infty,B_f\}$,
and process $P_2$ has the empty label $\{-/-\}$.
If process $P_1$ were allowed to send a message to $P_2$,
this action would leak $A$'s information via both message content
and the timing of the message's transmission,
and would leak $B$'s information (at a rate up to $f$)
via timing alone.
The system disallows this communication, therefore,
unless the processes hold and use the relevant capabilities
to adjust their labels before interacting.
In particular:
(a) $P_1$ must hold the capability $A^-$
and use it to remove its content tag $A$ before sending the message; and
(b) $P_1$ must hold and use a timing capability $B^-_f$ (or stronger) 
to declassify timing tag $B_f$ before sending the message.

\subsection{Controlling Timing Channels}

Timing labels and capabilities alone would not be useful
without mechanisms to control timing information flows.
This section briefly introduces two specific tools useful for this purpose:
{\em deterministic execution} and {\em pacing}.
The next section will illustrate how we might employ these tools
in practical systems.

\paragraph{Deterministic Execution:}

In general, a process whose label contains content tag $U$
must also have timing tag $U_\infty$,
because the process's flow of control---and hence execution time---%
can vary depending on the $U$-owned bits contained in its process state.
\com{
(We could envision special-purpose ``data-invariant''
execution models or processors that avoid flow control
and guarantee the same execution time on any input,
but such processors would be of limited use and are not our focus.)
}
The converse might also seem inevitable:
if a process has timing tag $U_f$ for any frequency $f$,
and the process reads the current time via \verb|gettimeofday()|, for example,
then the process's content subsequently depends on its execution timing,
hence the process must have content tag $U$.
Even if we disable system calls like \verb|gettimeofday()|,
conventional programming models---especially parallel, multithreaded models---%
enable processes and threads to depend on timing
in many implicit ways,
such as by measuring the relative execution speed of different threads.
One thread might simply remain in a tight loop 
incrementing a shared memory counter, for example,
which other threads read and use as a ``timer.''

System-enforced deterministic parallel execution,
as in Determinator~\cite{
	ford10efficient},
offers a tool to decouple a process's timing and content labels.
With system-enforced determinism,
the OS kernel can prevent unprivileged processes from exhibiting
{\em any} timing dependencies---%
even if the process maliciously attempts to introduce such dependencies---%
except via explicit inputs obtained through controlled channels.
In effect, deterministic processes cannot ``tell time''
except via explicit inputs
controlled by content labels.
System-enforced determinism
thus makes it ``safe'' for a process's content and timing labels to differ.
If a process's explicit inputs were derived from user $A$'s information,
but its execution timing
was also affected by $B$'s information at rate $f$,
we give the process the label $\{A/A_\infty,B_f\}$
rather than $\{A,B/A_\infty,B_f\}$,
safe in the knowledge that system-enforced determinism
prevents $B$'s ``timing domain'' information
from leaking into the process's ``content domain''
(its explicit register/memory state).

\paragraph{Pacing:}

Processes often interact with each other and with the external world
via queued messages or I/O,
and we leverage ``traffic shaping'' techniques common in networking
to limit the rate at which information might can across these queues
via timing channels.
We assume that we can {\em pace} the output of a message queue,
such that regardless of how messages build up in the queue,
the queue's output ``releases'' at most one message per tick
of a recurring timer, firing at some frequency $f$.
After each $1/f$-time period,
the queue's output releases exactly one message if the queue is non-empty,
and no message if the queue is empty.
Between clock ticks, the queue releases no information at all.
Ignoring information contained
in the {\em content} and {\em order} of queued messages---%
which we control via content labels---%
we see that a paced queue leaks at most one bit of timing information
per $1/f$-time period:
namely, whether or not the queue was empty at that particular timer tick.

If messages flowing into a paced queue
have a timing tag of $U_f'$ for $f' > f$ (including $f' = \infty$),
we can safely ``downgrade'' those timing tags to $U_f$
at the queue's output,
if the queue is paced at frequency $f$.
If messages with label $\{A/A_\infty,B_\infty\}$
flow into a pacer with frequency $f$, for example,
for example,
then messages at the queue's output
have label $\{A/A_f,B_f\}$.
While we for now can offer only an intuitive argument
for the correctness of this rate-limiting principle,
a formalized argument remains
for future TIFC model development.

\section{Using TIFC: Case Studies}

We now illustrate TIFC
with three simple examples,
in which two customers---%
Alice and Bob---%
each wish to perform a privacy-sensitive computation
on hardware managed by a trusted cloud provider.
Each customer desires strong assurance
that his data cannot leak to other customers
above a well-defined rate---%
even if his code is infected with malware
that attempts to leak his data via timing channels.
We make the simplifying assumption
that timing channels arise only from shared compute resources,
such as processor cores and the caches and functional units supporting them.
We neglect for now other sources of timing channels,
such as those arising from network communication paths
either within the cloud or between cloud and customers~\cite{
	ristenpart09cloud},
although this TIFC model may extend
to other channels as well.

\paragraph{Dedicated Hardware Scenario:}

\begin{figure*}[tbp]
\centering
\includegraphics[width=0.85\textwidth]{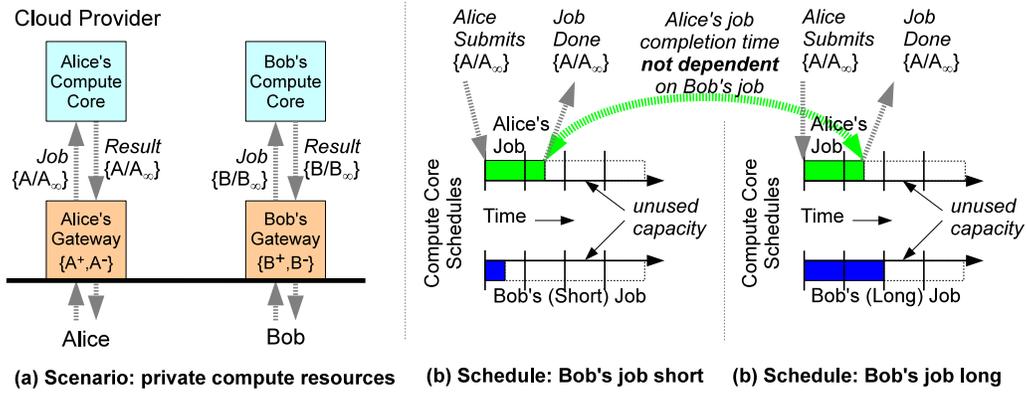}
\caption{Labeling Scenario:
	Private Per-Client Hardware Resources}
\label{fig-private}
\end{figure*}

The first example,
in Figure~\ref{fig-private}(a),
illustrates a ``base case'' scenario,
where the cloud provider controls timing channels
merely by imposing a fixed partitioning of hardware compute resources
between Alice and Bob.
Alice submits compute job requests
via a cloud gateway node that the provider dedicates exclusively to Alice,
and similarly for Bob.
Each customer's gateway forwards each job
to a compute core, on the same or another node,
that is also exclusive to the same customer.
The gateway nodes
attach TIFC labels to each incoming request,
and the provider's OS kernel or hypervisor
managing each compute core
uses these labels to prevent either customer's compute jobs
from leaking information to the other
via either the content or timing of messages within the cloud.

Figure~\ref{fig-private}(b) and (c)
illustrates the (intuitively trivial) reason
this example provides timing isolation,
by contrasting the system's timing when Bob submits a ``short'' job (b)
with the timing when Bob submits a ``long'' job (c).
Since Alice's job runs on a separate compute core from Bob's,
Alice's job completion time depends only on
the content of that job and Alice's prior jobs---%
information represented by the timing tag $A_\infty$---%
and is not ``tainted'' by any timing dependency on Bob's jobs.

\begin{figure*}[tbp]
\centering
\includegraphics[width=0.85\textwidth]{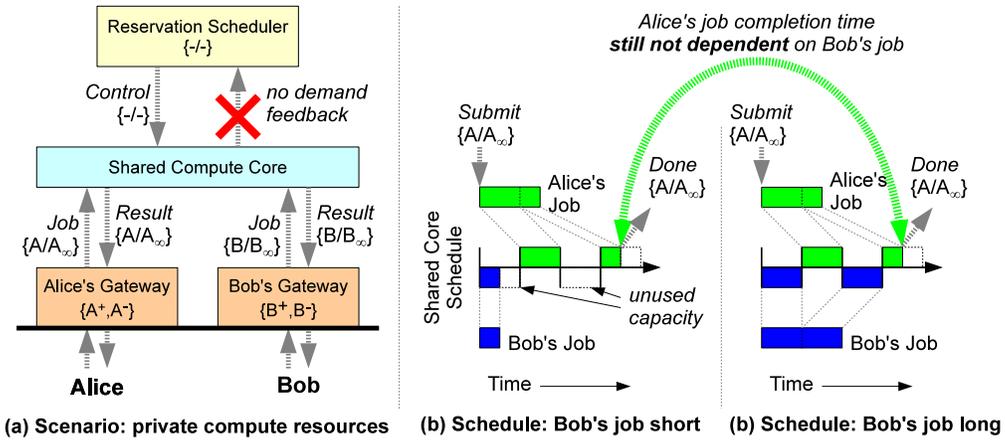}
\caption{Labeling Scenario: Shared Resource with Reservation-based Scheduling}
\label{fig-reserved}
\end{figure*}

\begin{figure*}[tbp]
\centering
\includegraphics[width=0.85\textwidth]{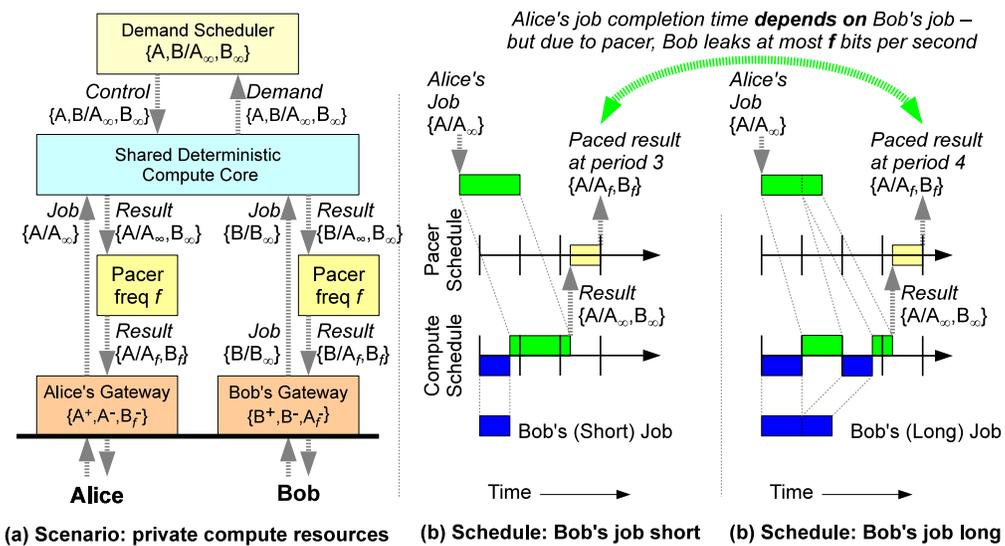}
\caption{Labeling Scenario: Shared Resource with Demand-driven Scheduling}
\label{fig-statmux}
\end{figure*}

\paragraph{Fixed-Reservation Timeslicing:}

Figure~\ref{fig-reserved}(a)
shows a less trivial example,
where a shared compute core
processes both Alice's and Bob's jobs
on a ``fixed reservation'' schedule that does {\em not} depend
on either Alice's or Bob's {\em demand} for the shared core.
The shared compute core
maintains and isolates the state of each customer's job
using standard process or virtual machine mechanisms.
The scheduling of these per-customer processors
onto the shared core, however,
is controlled by a separate entity we call the {\em reservation scheduler}.
The scheduler conceptually runs on its own dedicated CPU hardware,
and sends a message to the shared compute core
at the beginning of each timeslice
indicating which customer's job to run next.
The code implementing the scheduling policy
need not be trusted for information flow control purposes,
as long as trusted code attaches and checks TIFC labels appropriately.
In particular, the scheduler and the messages it sends have
the empty label $\{-/-\}$,
which allows the scheduler's messages to affect
the timing of Alice's and Bob's labeled jobs running on the shared core,
without adding any new ``taint.''

With its empty label, however,
the reservation scheduler cannot {\em receive} any messages
from the shared core
that might depend on either the content or timing of the customers' jobs.
TIFC enforcement prevents the scheduler
from obtaining feedback about whether either Alice's or Bob's processes
actually demand CPU time at any given moment,
forcing the scheduler to implement a ``demand-insensitive'' policy,
which isolates the timing of different customers' jobs sharing the core,
at the cost of wasting shared core capacity.
Figure~\ref{fig-reserved}(b) and (c)
shows execution schedules for the shared core
in the cases in which Bob's job is short or long, respectively,
illustrating why Alice's job completion time
depends only on Alice's information---%
hence the timing label of $A_\infty$---%
though Bob's job may have executed on the same core
during different (demand-independent) timeslices.

\paragraph{Statistical Multiplexing:}

The above scenarios
embody well-known timing channel control techniques~\cite{
	kemmerer83shared,wray91analysis},
to which TIFC merely adds an explicit,
analyzable and enforceable labeling model.
These standard techniques unfortunately
undermine the cloud {\em business model},
however,
by eliminating the cloud provider's ability
to obtain efficiencies of scale
through oversubscription and multiplexing~\cite{ford10determinating}.
Figure~\ref{fig-statmux} illustrates a final scenario
that {\em does} allow statistical multiplexing,
at the cost of a controlled-rate timing information leak.

As above,
this scenario includes a shared compute core
and a separate scheduler.
Instead of the empty (minimum) label, however,
we now give the scheduler a ``high'' (maximum) label
containing all customers' content and timing taints.
This label allows the scheduler to receive
demand information from the shared compute core,
and even to receive messages from customers' jobs themselves
containing explicit scheduling requests or ``performance hints.''
Since the scheduler's content label ($A,B$)
is higher than the content labels of either Alice's or Bob's jobs,
TIFC disallows the scheduler from sending messages
{\em to} Alice or Bob,
or otherwise affecting the {\em content} (process state) of their jobs.

The scheduler can send messages
to the shared compute core's trusted control logic, however,
to control
which customer's jobs run in a particular timeslice.
The shared core runs jobs deterministically,
ensuring that regardless of how the scheduler runs them,
each job's result content depends only on that job's input content
and not on execution timing.
The scheduler's control messages therefore ``taint'' all jobs
with the timing tags---but {\em not} the content tags---%
of all customers.
The results of Alice's job, for example,
has the label $\{A/A_\infty,B_\infty\}$,
indicating that the result content contains only Alice's information,
but the job's completion timing may also contain Bob's information.
Without additional measures,
this high timing label would prevent Alice's gateway
from sending Alice's job results back to Alice,
since the timing of these job completion messages
could leak Bob's information at an arbitrarily high rate.

To rate-limit this timing leak,
we assume that when requesting service from the cloud provider,
Alice and Bob agreed to allow timing information leaks
up to a specific rate $f$ fixed throughout this particular cloud.
To enforce this policy,
the cloud provider inserts a pacer on the path
of each customer's job results queue,
which releases the results of at most one queued job
at each frequency $f$ ``tick'' of a trusted provider-wide clock.
Since all customers allow
timing information leaks up to rate $f$,
each user's gateway node grants all other gateways
a timing declassification capability for rate $f$:
thus, Alice's and Bob's gateways can declassify each others' timing labels
up to rate $f$.
The TIFC rules thus allow Alice's job results to flow back to Alice
at up to $f$ jobs per second,
leaking at most $f$ bits per second of Bob's information.

Figure~\ref{fig-statmux}(b) and (c) compares two execution schedules
resulting from Bob's job being ``short'' and ``long,'' respectively.
Due to demand-sensitive multiplexing,
each job's completion time depends on the prior jobs of all users,
which may mix all users' information at arbitrary rate.
Alice's output pacer, however,
delays the release of each job's results to a unique clock tick boundary,
``scrubbing'' this timing channel down to the frequency $f$
at which the gateways can declassify the timing labels.

\section{Conclusion}
\label{sec-concl}

While TIFC may represent a promising approach
to hardening clouds against timing channels,
much work remains.
We are in the process of formalizing the model and security arguments,
and implementing it in an extension of Determinator~\cite{ford10efficient}
for experimental validation.

\begin{footnotesize}
\bibliography{os,net,soc}
\bibliographystyle{plain}
\end{footnotesize}

\end{document}